\documentclass{phb-proc4-auth}

\usepackage{graphicx}
\usepackage{amssymb}

\usepackage{bm}
\usepackage{amsmath}
\newcommand{\ff}[1]{{\bm #1}}

\begin{document}
\begin{frontmatter}

\journal{SCES '04}

\title{Disorder- and correlation-driven metal-insulator transitions}

\author{Matthias Balzer\thanksref{sfb}}
\author{and Michael Potthoff\thanksref{sfb}\corauthref{add}}

\address{Institut f\"ur Theoretische Physik und Astrophyik,
Universit\"at W\"urzburg, Germany}

\thanks[sfb]
{This work is supported by the DFG (Forschergruppe 538).}

\corauth[add]{Corresponding Author: 
Institut f\"ur Theoretische Physik und Astrophysik,
Universit\"at W\"urzburg,
Am Hubland,
97074 W\"urzburg, Germany.
Phone: +49-931-888-5724.
Fax: -5141.
Email: potthoff@physik.uni-wuerzburg.de}

\begin{abstract}
Metal-insulator transitions driven by disorder ($\Delta$) and/or by electron 
correlations ($U$) are investigated within the Anderson-Hubbard model with 
local binary-alloy disorder using a simple but consistent mean-field approach.
The $\Delta$--$U$ phase diagram is derived and discussed for $T=0$ and finite 
temperatures.
\end{abstract}


\end{frontmatter}

If spontaneous symmetry breaking is excluded, a system of electrons in a 
non-degenerate half-filled valence band may undergo a transition from a 
normal Fermi liquid to an insulator either due to Coulomb interaction 
or due to disorder.
Metal-insulator transitions (MIT) in the presence of strong electron 
correlations {\em and} disorder are not well understood -- even 
on the mean-field level. 
For the purely correlation-induced (Mott) MIT, the dynamical mean-field 
theory (DMFT) has uncovered a rather complex phase diagram \cite{GKKR96}.
The MIT to an (Anderson) insulator in case of non-interacting electrons 
and strong diagonal binary-alloy disorder can be described by the 
coherent-potential approximation (CPA) \cite{VKE68}.
While spatial correlations are neglected in both cases, the residual 
mean-field physics at low temperatures $T$ is non-trivial and relevant for 
three-dimensional transition-metal oxides, for example.
The combined problem can be studied within the half-filled ($n=1$) 
Anderson-Hubbard model (AHM):
$
H = - t \sum_{\langle ij \rangle, \sigma} c^\dagger_{i\sigma} c_{j\sigma}
+ \sum_{i\sigma} (\epsilon_{i} - \mu) n_{i\sigma}
+ U \sum_i  n_{i\uparrow} n_{i\downarrow} 
$.
Here the n.n.\ hopping is set to $t=1$, $U$ is the on-site interaction, 
$\mu = U/2$ the chemical potential, and $\epsilon_i = \pm \Delta/2$ with equal 
probabilities $x=1/2$ a random on-site energy at site $i$.
$\Delta$ measures the disorder strength.

\begin{figure}[t]
\centering
\includegraphics[width=0.8\columnwidth,clip=]{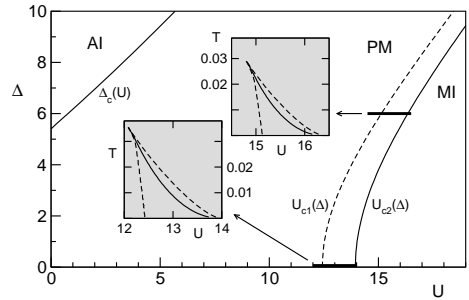}
\caption{
$U$-$\Delta$ phase diagram for $T=0$ (insets: $U$-$T$ phase diagram
for different $\Delta$). Energy scale: $t=1$. 
}
\label{fig:pd}
\end{figure}

``DMFT+CPA'' \cite{UJV95} can be regarded as the optimum mean-field approach 
to this model.
This, however, must be supplemented by stochastic \cite{UJV95} or 
renormalization-group techniques \cite{BHV03} or by further approximations, 
e.g.\ weak-coupling perturbation theory \cite{LCMH01}.
While parts of the phase diagram in the $U$-$\Delta$-$T$ space are
known \cite{UJV95,LCMH01}, a comprehensive study is still missing.

Here we employ the self-energy-functional approach (SFA) \cite{sfa} and 
the $n_{\rm s}=2$-site dynamical-impurity approximation (2S-DIA) which 
nicely reproduces the phase diagram for $\Delta = 0$ \cite{sfa}.
In case of disorder, a proper generalization of the formalism has to be 
applied \cite{PB04}.
Within the generalized framework, the 2S-DIA can be regarded as a strongly 
simplified but consistent DMFT-CPA approach.
In the limit $n_{\rm s} \to \infty$ one recovers the DMFT for $\Delta=0$, 
the CPA for $U=0$ and the DMFT-CPA for $U,\Delta\ne 0$.

Operationally, the Green's function $\ff G'$ of a single-impurity Anderson 
model $H'$ with two sites and impurity on-site energies $\epsilon = \pm \Delta$ 
is obtained by exact diagonalization and averaged, 
$\ff \Gamma' = \langle \ff G' \rangle$, to get the configuration-independent 
self-energy $\ff S' \equiv {\ff G'_0}^{-1} - {\ff \Gamma'}^{-1}$ where ${\ff G'_0}$
is the free ($U,\Delta=0$) Green's function.
$\ff S'=\ff S(\ff t')$ depends on the one-particle parameters of $H'$ and
is used as a trial self-energy in a general variational principle,
$\delta \Omega[\ff S] = 0$, which gives the exact averaged grand potential
of the AHM at the physical $\ff S$.
On the subspace given by $\ff S(\ff t')$, the functional can be evaluated
rigorously (see \cite{PB04}).
We consider the paramagnetic phase of the AHM on a three-dimensional s.c.\ 
lattice consisting of $10^3$ sites.
Phase boundaries are obtained from the 
resulting $\Omega$ as a function of $\Delta$, $U$ and $T$.
The averaged interacting local density of states (DOS) of the AHM can be 
calculated via
$\rho(\omega)=-\mbox{Im} \, \Gamma_{ii}(\omega+i\eta)/\pi$ 
and $\ff \Gamma = (\ff G_0^{-1} - \ff S')^{-1}$ where $\ff G_0$ is the free 
($U,\Delta=0$) {\em lattice} Green's function.

Three different phases are identified at $T=0$
(see Fig.\ \ref{fig:pd}): a paramagnetic metallic phase (PM), a Mott 
insulator (MI), and an Anderson insulator (AI).
For any disorder strength $\Delta$, we find the AI at weak $U$ (and 
$\Delta \ge \Delta_{\rm c}(U)$) to be well separated from the MI at 
strong $U$ by the PM in between.
For $\Delta=0$ the critical interaction for the Mott MIT is found to be
$U_{\rm c} = 13.9 \approx 1.16W$ (with $W=12$ the free band width) while 
$\Delta_{\rm c}=5.4=0.46W$ for the MIT at $U=0$. 
This agrees well with full DMFT and CPA estimates, respectively 
\cite{GKKR96,VKE68,sfa}.
For $U_{\rm c1} \le U \le U_{\rm c2}$, a coexistence of the stable PM 
phase with the metastable MI phase is observed ($U_{\rm c1} = 12.4$).
This scenario for the Mott MIT is well known for $\Delta=0$ \cite{GKKR96}
and is shown here to survive for any finite disorder strength with a
$\Delta$ dependent coexistence region 
$U_{\rm c1}(\Delta) \le U \le U_{\rm c2}(\Delta)$
and $U_{\rm c}(\Delta) = U_{\rm c2}(\Delta)$.
A discontinuous Mott MIT with 
$U_{\rm c1}(\Delta) \le U_c(\Delta) \le U_{\rm c2}(\Delta)$
is found for finite temperatures $0 < T \le T_{\rm c}(\Delta)$.
For $T \ge T_{\rm c}(\Delta)$ there is a smooth crossover only.
For $\Delta \to \infty$, the critical interactions 
approach a linear dependence, $U_{\rm c1,2}(\Delta) \to \Delta + \mbox{const}_{1,2}$ 
while $T_{\rm c}(\Delta) \to T_{\rm c}$ saturates.

\begin{figure}[t]
\centering
\includegraphics[width=0.95\columnwidth,clip=]{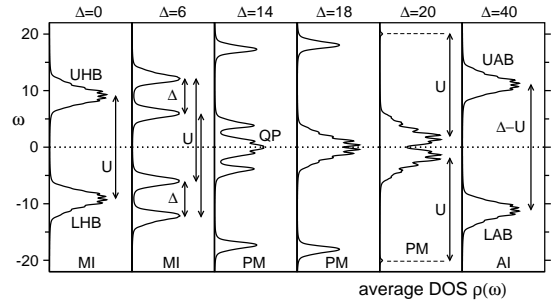}
\caption{
Average density of states for $U=18$ and different $\Delta$ and 
Lorentzian broadening with $\eta = 0.25$.
}
\label{fig:dos}
\end{figure}

The topology of the phase diagram can be understood by looking at the DOS, 
see Fig.\ \ref{fig:dos}.
Characteristic for the MI at $\Delta=0$ and $U > U_{\rm c}$ is the insulating gap 
between the lower and upper Hubbard band (LHB, UHB).
For finite $\Delta$ the gap decreases due to a broadening and, eventually, a splitting 
of each of the Hubbard bands (Fig.\ \ref{fig:dos}, $\Delta=6$).
The closure of the gap is preempted by the occurrence of a quasi-particle peak (QP)
at $\omega=0$ which marks the transition to the PM (see $\Delta=14$).
Apart from the QP, the spectrum can be understood as being composed of two Hubbard 
bands at $\omega \approx \pm \Delta/2 \pm U/2$ for each of the two atomic configurations 
$\epsilon = \pm \Delta/2$.
This explains the strong spectral-weight transfer when increasing $\Delta=18$ to 
$\Delta=20$:
As the $\epsilon = +\Delta/2$-DOS ($\epsilon = -\Delta/2$-DOS) becomes almost completely 
unoccupied (occupied), the weight of the UHB (LHB) must disappear.
Finally, a further increase of $\Delta$ induces a splitting into an upper and lower 
alloy band (UAB, LAB) and a MIT to the AI.

Recently, Byczuk et al \cite{BHV03} have shown that DMFT+CPA predicts the AHM 
to exhibit a Mott MIT also for fillings $n \ne 1$ if $x=n$.
Similar to the presently considered case $n=1=2x$, a sharp QP at $\omega=0$ (even 
for strong disorder) as well as a coexistence of the PM and the MI is found close 
to the MIT.
We like to point out that the phase diagram for $n = x \ne 1$ \cite{BHV03} can be 
understood by an analysis of the DOS completely analogous to the $n=1=2x$ case
discussed above - although its topology is quite different.

Concluding, we have proposed a mean-field scenario for the 
MIT in the AHM at half-filling $n=1=2x$ on the basis of a simplified DMFT+CPA approach.
The phase diagram can be understood by a quasi-atomic
interpretation of the DOS in combination with the Mott MIT scenario of the pure system.
This should be contrasted with full DMFT+CPA calculations in the future which may also 
clarify the importance of disorder scattering due to a finite self-energy 
$\mbox{Im} \ff S(\omega=0)$ which has been neglected here.

\end{document}